\documentstyle[aps,multicol,epsfig]{revtex}

\newcommand{\simle}
{\raisebox{-0.75ex}[-1.5ex]{$\;\stackrel{<}{\sim}\;$}}
\newcommand{\simge}
{\raisebox{-0.75ex}[-1.5ex]{$\;\stackrel{>}{\sim}\;$}}
\def\d{\partial}
\def\s{{\sigma}}
\def\e{{\epsilon}}
\def\k{{ {\bf k} }}

\def\q{{ {\bf q} }}

\def\w{{\omega}}

\def\i{{ {\rm i} }}

\begin{document}

\def\runtitle{
Theory of Thermoelectric Power in High-$T_{\rm c}$ Superconductors
}
\def\runauthor
 {Hiroshi {\sc Kontani}}

\title{
Theory of Thermoelectric Power in High-$T_{\rm c}$ Superconductors
}

\author{
Hiroshi {\sc Kontani}
}

\address{
Department of Physics, Saitama University,
255 Shimo-Okubo, Urawa-city, 338-8570, Japan
}

\date{\today}

\maketitle

\begin{abstract}
We present a microscopic theory for 
the thermoelectric power (TEP) in high-$T_{\rm c}$ cuprates.
Based on the general expression for the TEP,
we perform the calculation of the TEP for a
square lattice Hubbard model
including all the vertex corrections necessary to satisfy
the conservation laws.
In the present study,
characteristic anomalous temperature and doping dependences
of the TEP in high-$T_{\rm c}$ cuprates,
which have been a long-standing problem of high-$T_{\rm c}$ cuprates,
are well reproduced for both hole- and electron-doped systems,
except for the heavily under-doped case.
According to the present analysis,
the strong momentum and energy dependences of the self-energy 
due to the strong antiferromagnetic fluctuations
play an essential role in reproducing experimental
anomalies of the TEP.
\end{abstract}


\sloppy

\begin{multicols}{2}

The thermotransport phenomena in high-$T_{\rm c}$ cuprates
exhibit various anomalous behaviors.
In particular, the thermoelectric power (TEP),
expressed by the Seebeck coefficient $S$,
is known to show characteristic temperature and
doping dependences above $T_{\rm c}$,
namely
{\bf (i)} $dS/dT<0$ in hole-doped compounds such as 
YBa$_2$Cu$_3$O$_{7-x}$ (YBCO) or 
La$_{2-x}$Sr$_x$CuO$_4$ (LSCO), and
$S$ is positive in the under-doped systems at room temperatures
 \cite{Cooper,Sato}.
{\bf (ii)} $S$ is negative in electron-doped compounds like 
Nd$_{2-x}$Ce$_x$CuO$_4$ (NCCO) and 
LaPr$_{1-x}$Ce$_x$CuO$_4$ (PCCO), and 
$dS/dT>0$ in the under-doped compound at higher temperatures
$T\simge200$~K
 \cite{Sato,Fournier}.
In both cases, $|S|$ increases drastically as the doping decreases.
Thus, a conventional Fermi liquid type behavior,
$S\propto T$, is totally violated in high-$T_{\rm c}$ cuprates
for a wide range of temperatures.
In particular,
a qualitative particle-hole symmetric behavior of the TEP, i.e.,
$S_{\rm LSCO} \sim -S_{\rm NCCO}$ for the same carrier doping 
weight $x$ as was reported in ref.
 \cite{Sato},
is very mysterious because both LSCO (YBCO) and NCCO
have similar large hole-like Fermi surfaces according to 
angle-resolved photoemission spectroscopic (ARPES) studies
 \cite{ARPES-NCCO,ARPES-NCCO2}.


From an academic point of view,
the TEP is a unique and important phenomenon
in that it sensitively reflects the properties of the quasiparticles
away from the Fermi surface, whereas electronic transport phenomena
are caused by the quasiparticles on the Fermi surface.
Thus, 
the characteristic non-Fermi liquid behavior of the TEP 
in high-$T_{\rm c}$ cuprates suggests that 
the excited states of quasiparticles in high-$T_{\rm c}$ cuprates
are significantly anomalous.
In this respect, the theoretical study of the TEP is 
quite important.

Up to now, several theoretical studies on the TEP 
for models with strong Coulomb interactions
have been performed based on the dynamical mean field theory (DMFT)
 \cite{d-inf1,d-inf2}.
These studies found that the energy dependence of the relaxation time
is important for the TEP in strongly correlated systems.
However, 
in high-$T_{\rm c}$ cuprates,
it is known that 
the momentum dependence of the relaxation time is also prominent, 
and the vertex correction (VC) in terms of 
the current plays an essential role
for transport phenomena.
These features are totally dropped in DMFT.
Thus, for the present purpose,
we have to develop a theory for the TEP
by taking the momentum dependence into account.

In this letter,
we study the TEP in high-$T_{\rm c}$ cuprates
on the basis of the antiferromagnetic (AF) fluctuation theory.
We take the momentum and energy dependences of the relaxation
time into account, and include all of the VC's
required by the Ward identity, i.e.,
the conserving approximation
 \cite{Baym}.
Our study reproduces the main futures of the experimental TEP
for both hole- and electron-doped compounds.
According to the present analysis,
the approximate particle-hole symmetry for the TEP is
caused by the "alternation of cold spots",
as is the case with the Hall coefficient 
 \cite{Kontani,Kanki}


In the present study,
we calculate the self-energy $\Sigma_\k(\e)$ using 
the fluctuation-exchange (FLEX) approximation
which is a kind of self-consistent spin-fluctuation theory
 \cite{Bickers},
because it works well for high-$T_{\rm c}$ cuprates
for $|1-n|\simge0.1$ above the spin pseudo-gap temperature
 \cite{Dahm}. 
Here,
we study the square lattice tight-binding model 
whose dispersion is given by
$\e_\k^0 = 2t(\cos k_x + \cos k_y) + 4t' \cos k_x \cos k_y
 + 2t''(\cos 2k_x + \cos 2k_y)$,
where $t$, $t'$ and  $t''$ denote the nearest-, next-nearest- 
and third-nearest-neighbor hopping integrals.
The parameters are chosen as 
$(t,t',t'')=(-1.0,0.17,-0.2)$ for YBCO and NCCO
 \cite{Kontani},
and $(-1.0,0.15,-0.05)$ for LSCO
 \cite{MR-HTSC}, respectively.

First, we discuss the linear response theory for the TEP.
In 1964, Luttinger derived a general expression for the TEP
based on the linear response theory
 \cite{Luttinger}.
Although his original formula was given by a complicated 
three-particle correlation function, 
Jonson and Mahan later obtained a much simplified formula 
for a system with impurities and adiabatic phonons 
 \cite{Mahan}.
In the same way, the formula for the TEP
of a Hubbard model with on-site Coulomb
interaction $U$ is derived as follows
 \cite{Future}: 
\begin{eqnarray}
 S &=& -L_{xx}/eT\s_{xx},
  \\
 L_{xx} &=& \left. {\Phi(\w+\i0)}/{\i\w} \right|_{\w\rightarrow0},
\end{eqnarray}
where $\Phi(\w+\i0)$ is given by the analytic continuation 
of the following function:
\begin{eqnarray}
 \Phi(\w_\lambda) =
  \int_0^\beta d\tau e^{-\w_\lambda \tau} \w_\lambda
  \langle T_\tau j_x(0) j_x(\tau) \rangle,
\end{eqnarray}
where ${\vec j}_\k$ is the current operator, 
$T_\tau$ is a $\tau$ ordering operator, 
and $\w_\lambda = 2\pi\i T\lambda$.
Here, $\s_{xx}$ is the conductivity and
$-e$ ($e>0$) is the charge of an electron.
Recently, we performed the analytic continuation
{\it exactly} as for the most singular terms 
with respect to the inverse of the damping rate of
the quasiparticles
 \cite{Future}.
The result is
\begin{eqnarray}
 L_{xx} &=& 
 \sum_\k \int \frac{d\e}{\pi}
 \left(-\frac{\d f}{\d\e} \right) \e v_{\k x}(\e)
 \Bigl[ \ |G_\k(\e)|^2 J_{\k x}(\e) 
  \nonumber \\
 & &-  
 {\rm Re} \left\{G_\k^2(\e) \right\} v_{\k x}(\e)  \ \Bigr] ,
  \label{eqn:Lxx} \\
  J_{\k x}(\e) &=& v_{\k x}(\e) + \sum_{\k'} 
  \nonumber \\
 & &\int \frac{d\e'}{4\pi\i}  {\cal T}_{22}^I(\k\e|\k'\e')
 \left| G_{\k'}(\e')\right|^2 J_{\k' x}(\e'),
  \label{eqn:J} 
\end{eqnarray}
where $v_{\k x}(\e) = (\d/\d k_x)(\e_\k^0 + {\rm Re}\Sigma_\k(\e))$,
$f(\e)= (1+\exp((\e-\mu)/T))^{-1}$, and
$G_\k(\e)$ is the Green function.
Here, $J_{\k x}(\e)$ is the total current with the VC from the
irreducible four-point vertex ${\cal T}_{22}^I$,
which was introduced by Eliashberg in 
 ref.~\cite{Eliashberg}
for the first time.
The derivation of eq.~(\ref{eqn:Lxx})
beased on the Fermi liquid theory will be given elsewhere
 \cite{Future}.

In calculating a transport coefficient,
it is known that the conservation laws should be satisfied
to avoid unphysical results
 \cite{Yamada}.
For this purpose,
the Ward identity between 
${\cal T}_{22}^I$ in eq.~(\ref{eqn:J})
and the self-energy $\Sigma_\k(\e)$
has to be taken into account
 \cite{Yamada,Baym}.
Note that ${\cal T}_{22}$ within the FLEX approximation
is given in ref.~\cite{Kontani}.
We stress that 
the relaxation time approximation cannot reproduce the 
{\it seemingly} non-Fermi liquid behaviors of $R_{\rm H}$ 
and the magnetoresistance ($\Delta\rho/\rho$):
These long-standing mysteries 
of the magnetotransport phenomena were recently solved 
by taking account of the VC for ${\vec J}_\k$
in refs.~\cite{Kontani} and \cite{Kanki}for $R_{\rm H}$ 
and in refs.~\cite{MR-HTSC} and \cite{MR-formula}
for $\Delta\rho/\rho$,
from the standpoint of the nearly AF Fermi liquid.
Moreover, anomalous temperature and pressure dependences of 
$R_{\rm H}$ in $\kappa$-BEDT-TTF organic superconductors
are well reproduced according to the same mechanism
 \cite{BEDT-Hall}.
Thus, it is important to determine whether the 
spin-fluctuation theory with satisfying the conservation laws
is also successful for the thermotransport phenomena.

As for the conductivity, 
we calculate $\s_{xx}$ based on the conserving approximation
as explained in ref.~\cite{Kontani}.
According to the FLEX approximation,
$\s_{xx} \propto 1/\gamma_{\rm cold}$ 
because the VC gives only a qualitative correction for $\s_{xx}$.
As a result,
the famous $T$-linear resistance
in high-$T_{\rm c}$ cuprates is reproduced for a
wider range of temperatures
 \cite{Kontani}.

Before calculating the TEP,
we consider the behavior of eq.~(\ref{eqn:Lxx}) 
by using the quasiparticle representation of the Green function,
$G_\k(\w+\i\delta) = z_\k/(\w-\e_\k^\ast +\i\gamma_\k^\ast)$.
Here, $\e_\k^\ast = z_\k(\e_\k^0+{\rm Re}\Sigma_\k(\e_\k^\ast)-\mu)$,
$\gamma_\k^\ast= -z_\k{\rm Im}\Sigma_\k(\e_\k^\ast+\i0) >0$,
and $z_\k= (1-\d\Sigma_\k(\e)/\d\e)^{-1}$ is the
renormalization factor.
Then, eq.~(\ref{eqn:Lxx}) is simplified as 
\begin{eqnarray}
 L_{xx} &=& \int_{\rm FS} \frac{dk_\parallel}{(2\pi)^2}
  \int dk_\perp
  z_\k \left(-\frac{\d f}{\d\e}\right)_{\e=\e_\k^\ast} 
  \frac{\e_\k^\ast}{\gamma_\k(\e_\k^\ast)}
   \nonumber \\
& &\times
 v_{\k x}(\e_\k^\ast) J_{\k x}(\e_\k^\ast) ,
 \label{eqn:Lxx-lowT}
\end{eqnarray}
where $k_\parallel$ [$k_\perp$]
is the momentum along [perpendicular to] the Fermi surface.
At sufficiently low temperatures, 
eq.~(\ref{eqn:Lxx-lowT}) becomes
\begin{eqnarray}
L_{xx} &=& \frac{(\pi T)^2}{6} \int_{\rm FS} 
 \frac{dk_\parallel}{(2\pi)^2}
 \frac1{z_\k|v_\k(\e_\k^\ast)|^2}
  \nonumber \\
& &\times
 \frac{\d}{\d k_\perp}
 \left(\frac{v_{\k x}(\e_\k^\ast) J_{\k x}(\e_\k^\ast)}
 {\gamma_\k(\e_\k^\ast)}\right) .
  \label{eqn:Lxx-lowT2}
\end{eqnarray}
Thus, $L_{xx}$ given by eq.~(\ref{eqn:Lxx-lowT2})
is enhanced by $z_\k^{-1}$.

When the temperature dependences of $\gamma_\k(\e_\k^\ast)$, 
$v_{\k x}(\e_\k^\ast)$ and $J_{\k x}(\e_\k^\ast)$ 
are negligible like in a conventional Fermi liquid,
then $L_{xx}\propto T^2 /\gamma$ and $S\propto T$ 
are obtained.
However, the temperature dependences of these functions 
are usually large at higher temperatures in strongly correlated systems.
For example, in heavy Fermion systems,
a huge $\e$ dependence 
of $\gamma(\e)$
due to the Kondo resonance 
causes a prominent non-Fermi liquid behavior on $S$
around the Kondo temperature
 \cite{d-inf1,d-inf2}.
In the present study,
we find that the temperature dependences of 
the anisotropy of $\gamma_\k(\e_\k^\ast)$ 
in eqs.~(\ref{eqn:Lxx-lowT}) or (\ref{eqn:Lxx-lowT2}) 
are primarily responsible for highly enhanced TEP in 
(under-doped) high-$T_{\rm c}$ cuprates.
Thus, we analyze the behavior of $\gamma_\k(\e_\k^\ast)$ below.

\begin{figure}
\begin{center}
\epsfig{file=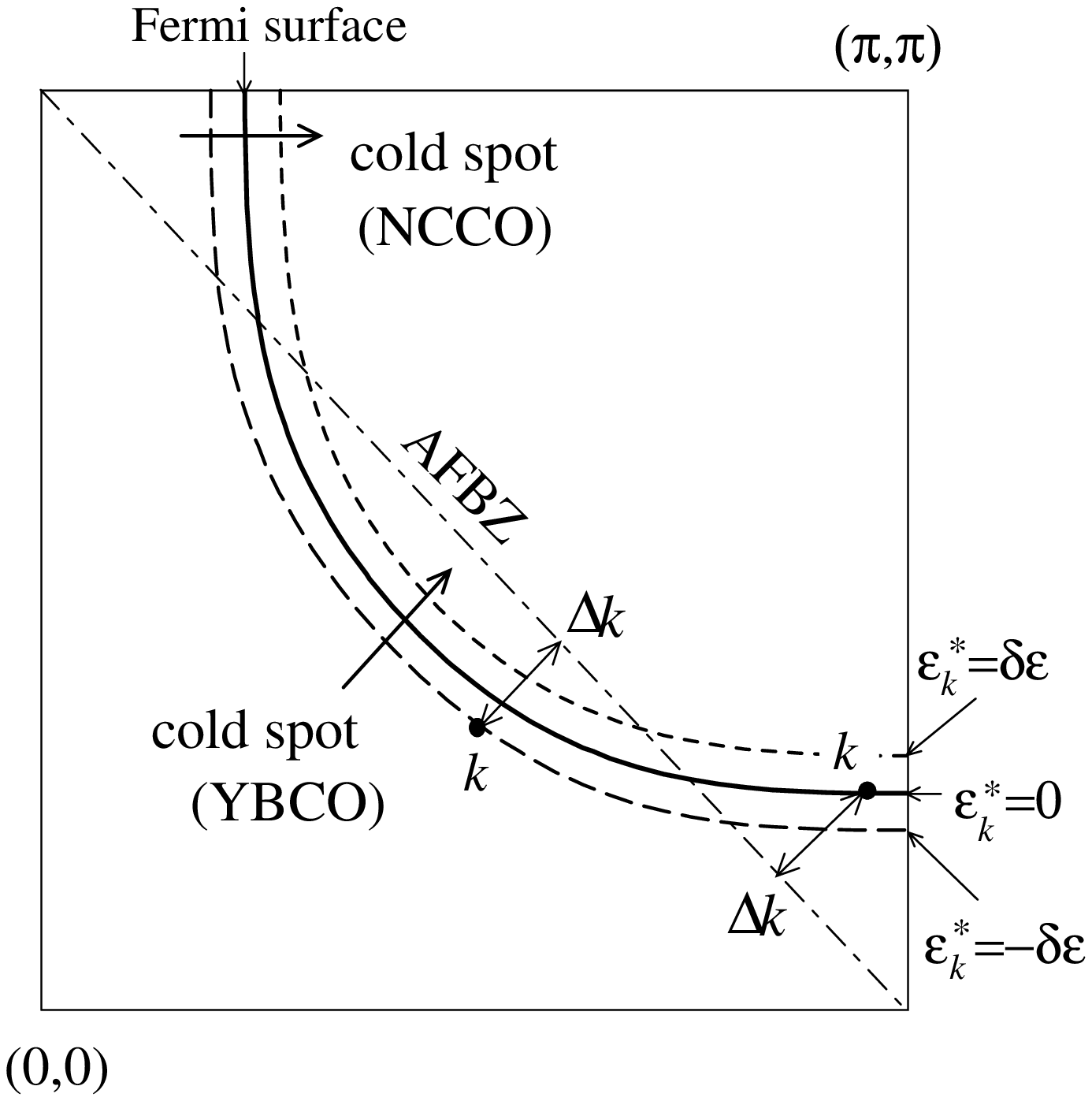,width=6.5cm}
\end{center}
\caption{
Fermi surface in YBCO/LSCO.
AFBZ stands for the AF Brillouin zone boundary.
The cold spot, at which Im$\Sigma_{\k}(0)$ 
takes the smallest value on the Fermi surface,
is located near $(\pi/2,\pi/2)$ [$(0,\pi)$]
in YBCO [NCCO].
}
  \label{fig:FS-S}
\end{figure}
Figure \ref{fig:FS-S}
shows the Fermi surface without interaction,
together with the contour given by
$\e_\k^\ast = \pm \delta\e$ ($\delta\e>0$).
According to the ARPES measurement,
the position of the cold spot, where $\gamma_\k$ takes
the smallest value, is near $(\pi/2,\pi/2)$ in YBCO
and near $(\pi,0)$ in NCCO.
This "cold spot alternation"
was first predicted using the FLEX approximation in ref. 
 \cite{Kontani} prior to the ARPES measurement 
 \cite{ARPES-NCCO,ARPES-NCCO2}.
According to the analysis in ref.
 \cite{Kontani},
this finding leads to the opposite sign of the Hall coefficient
in hole- and electron-doped compound, although they have similar
hole-like Fermi surfaces.
Later in this article,
this finding also explains the opposite sign of the TEP
in hole- and electron-doped compound.

\begin{figure}
\begin{center}
\begin{minipage}{0.45\linewidth}
\epsfig{file=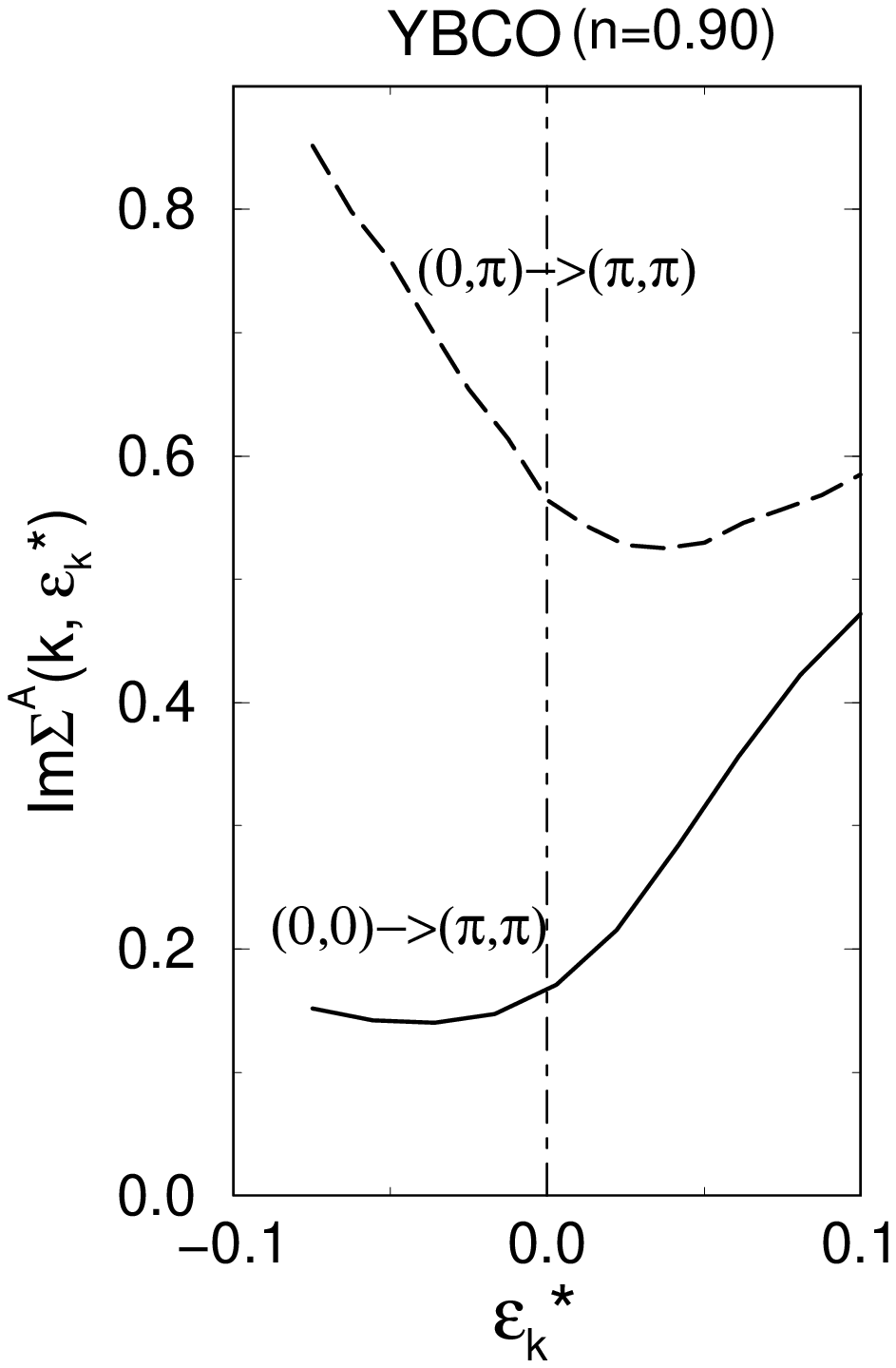,width=3.2cm}
\end{minipage}
\begin{minipage}{0.45\linewidth}
\epsfig{file=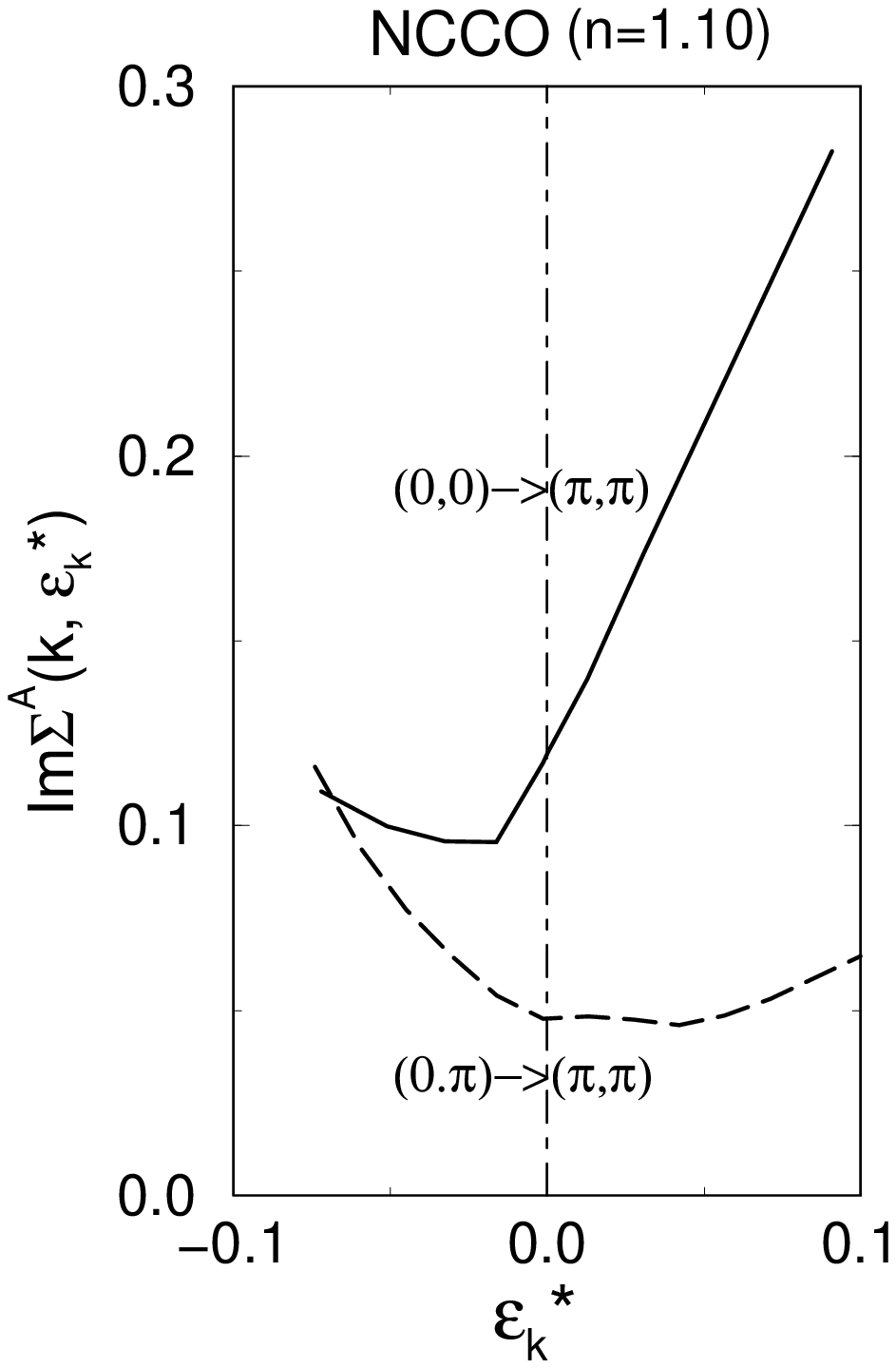,width=3.4cm}
\end{minipage}
\end{center}
\caption{
Obtained relation between
$\gamma_\k \equiv {\rm Im}\Sigma_\k^A(\e_\k^\ast)$
and $\e_\k^\ast$ at $T=0.02$.
We see that 
$\d\gamma_\k(\e_\k^\ast)/\d k_\perp >0$ $(<0)$ around the cold spot
for YBCO (NCCO), where $k_\perp$ is normal to the Fermi surface.
}
  \label{fig:del-Ek}
\end{figure}
Next, we examine the $(\k,\e)$-dependence of the 
self-energy $\Sigma_\k(\e)$ given by the FLEX approximation.
Here we use $U=8$ for YBCO and $U=6$ for NCCO.
Figure \ref{fig:del-Ek}
shows the obtained $\gamma_\k(\e_\k^\ast)$ vs $\e_\k^\ast$
near the Fermi surface, 
along $(0,0)\rightarrow(\pi,\pi)$ 
and $(0,\pi)\rightarrow(\pi,\pi)$.
Here, $\e_\k^\ast$ is the solution of 
${\rm Re}\{1/G_\k(\e_\k^\ast)\} = 0$.
We stress that $\gamma_\k(\e_\k^\ast)$ is highly asymmetric,
and $\d\gamma_\k(\e_\k^\ast)/\d k_\perp >0$ [$<0$]
at the cold spot in YBCO [NCCO],
where $k_\perp$ is the momentum normal to the Fermi surface.
(Note that $\gamma_\k(\e_\k^\ast) \propto \{ \e_\k^\ast \}^n$
and $n\sim2$ if the self-energy is $k$-independent.)
This finding naturally explains why the sign of ${S}$ 
in a hole-doped compound and that in an electron-doped compound
are different,
according to eq.~(\ref{eqn:Lxx-lowT2}) for $L_{xx}$.

Here, we analyze the origin of the asymmetric
behavior of $\gamma_\k(\e_\k^\ast)$
in terms of the nearly AF Fermi liquid:
In high-$T_{\rm c}$ cuprates,
the spin propagator is well expressed in the following 
functional form:
\begin{eqnarray}
 \chi_\q(\w) = \chi_Q/(1+\xi^2(\q-{\bf Q})^2 -\i\w/\w_{\rm sf}),
\end{eqnarray}
where ${\bf Q}= (\pi,\pi)$, $\xi$ is the AF correlation length, and
$\chi_Q \propto \w_{\rm sf}^{-1} \propto \xi^2$.
According to the standard AF fluctuation theory,
$\xi^2 \propto T^{-1}$ in two-dimensional systems
 \cite{Dahm,SCR}.
If we assume $\w_{\rm sf} \gg T$ for simplicity,  
then $\gamma_\k(\e)= {\rm Im}\Sigma_\k(\e-\i 0)$
around the cold spot is approximately given by
\begin{eqnarray}
 \gamma_\k(\e) \propto \frac1{|v_\k|}
 \frac{\xi^3 (T^2 + (\e/\pi)^2)}{ [ \ 1+\xi^2(2\Delta k)^2 \ ]^{3/2} } ,
   \label{eqn:gamma}
\end{eqnarray}
within the one-loop approximation
 \cite{Pines}.
Here, $\Delta k$ is the distance between $\k$ and the 
antiferromagnetic Brillouin zone (AFBZ) boundary, as shown
in Fig.~\ref{fig:FS-S}.
Equation (\ref{eqn:gamma}) gives an analytical explanation
for the numerical results shown in Fig.~\ref{fig:del-Ek}.
(According to the FLEX approximation,
$\xi \simle 1/\Delta k$ for $n\le 0.9$
 \cite{Kontani}.)
By considering Fig.~\ref{fig:FS-S}, 
eq.~(\ref{eqn:gamma}) directly indicate that 
{\bf (i)} the inside region of the AFBZ provides the positive
contribution for $S$ because $\d\gamma_\k(\e_\k^\ast)/\d k_\perp >0$
in the presence of strong AF fluctuations.
At the same time,
{\bf (ii)} the outside of the AFBZ provides the negative contribution
because $\d\gamma_\k(\e_\k^\ast)/\d k_\perp <0$.

\begin{figure}
\begin{center}
\epsfig{file=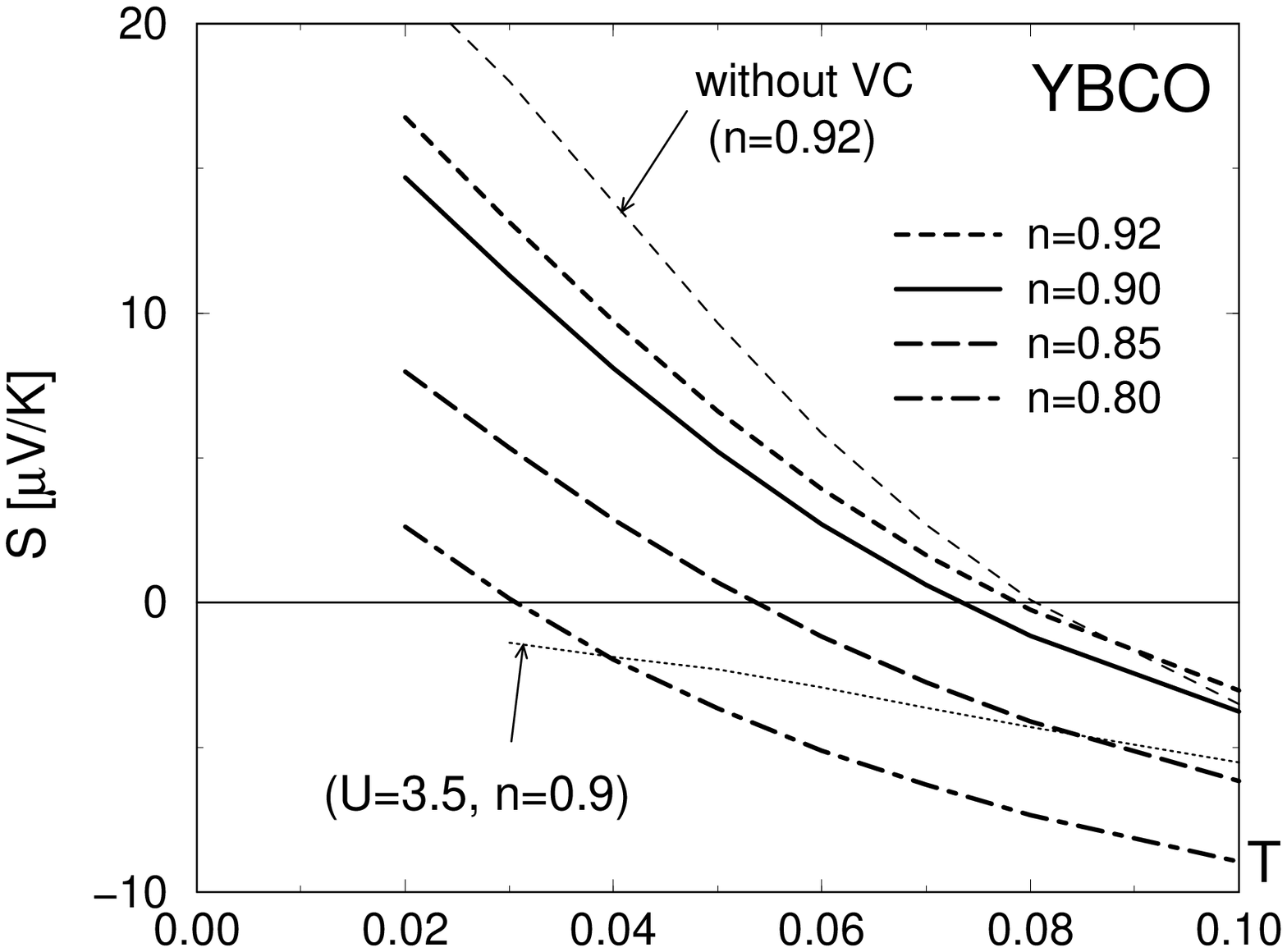,width=6.5cm}
\epsfig{file=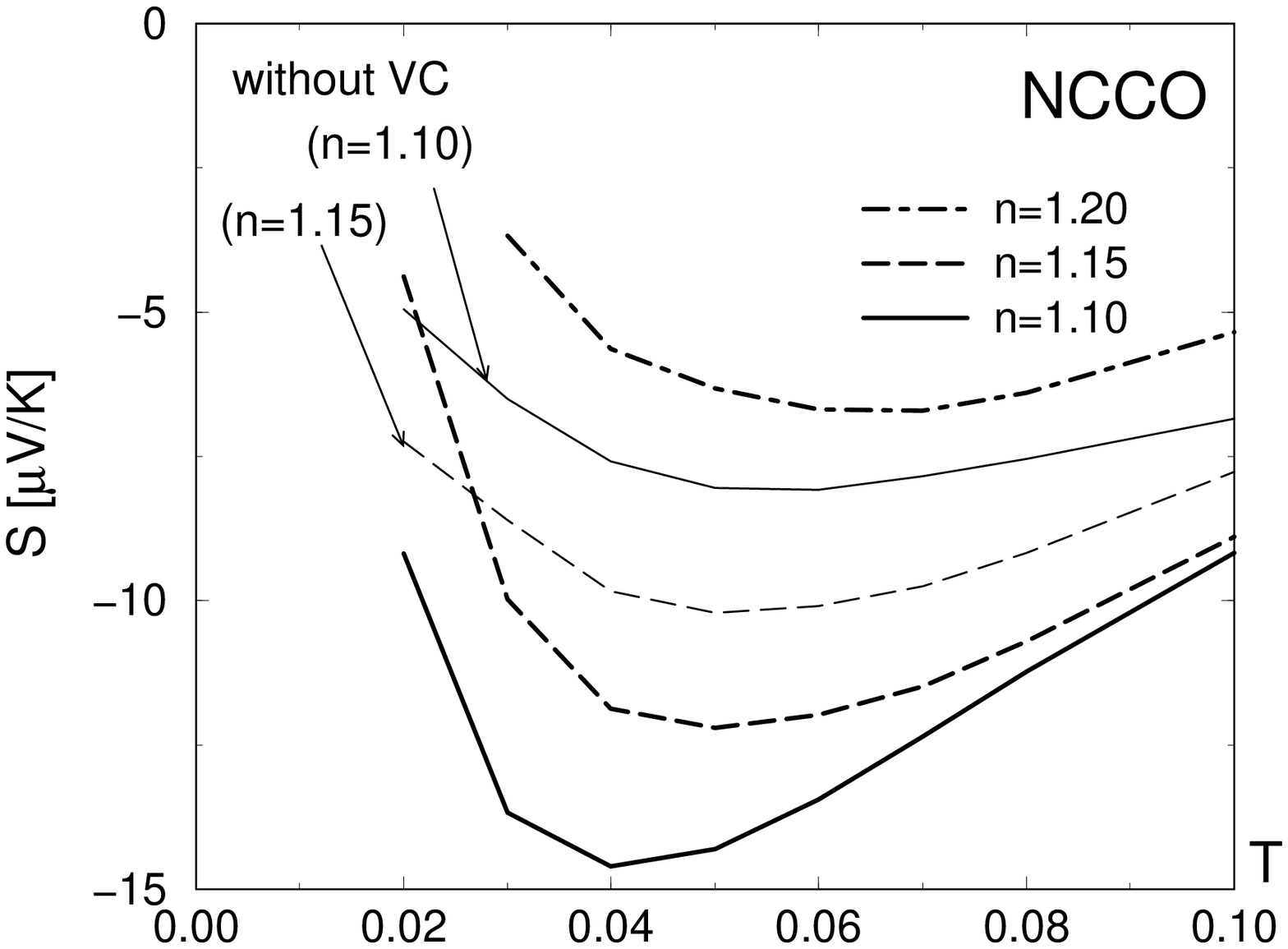,width=6.5cm}
\epsfig{file=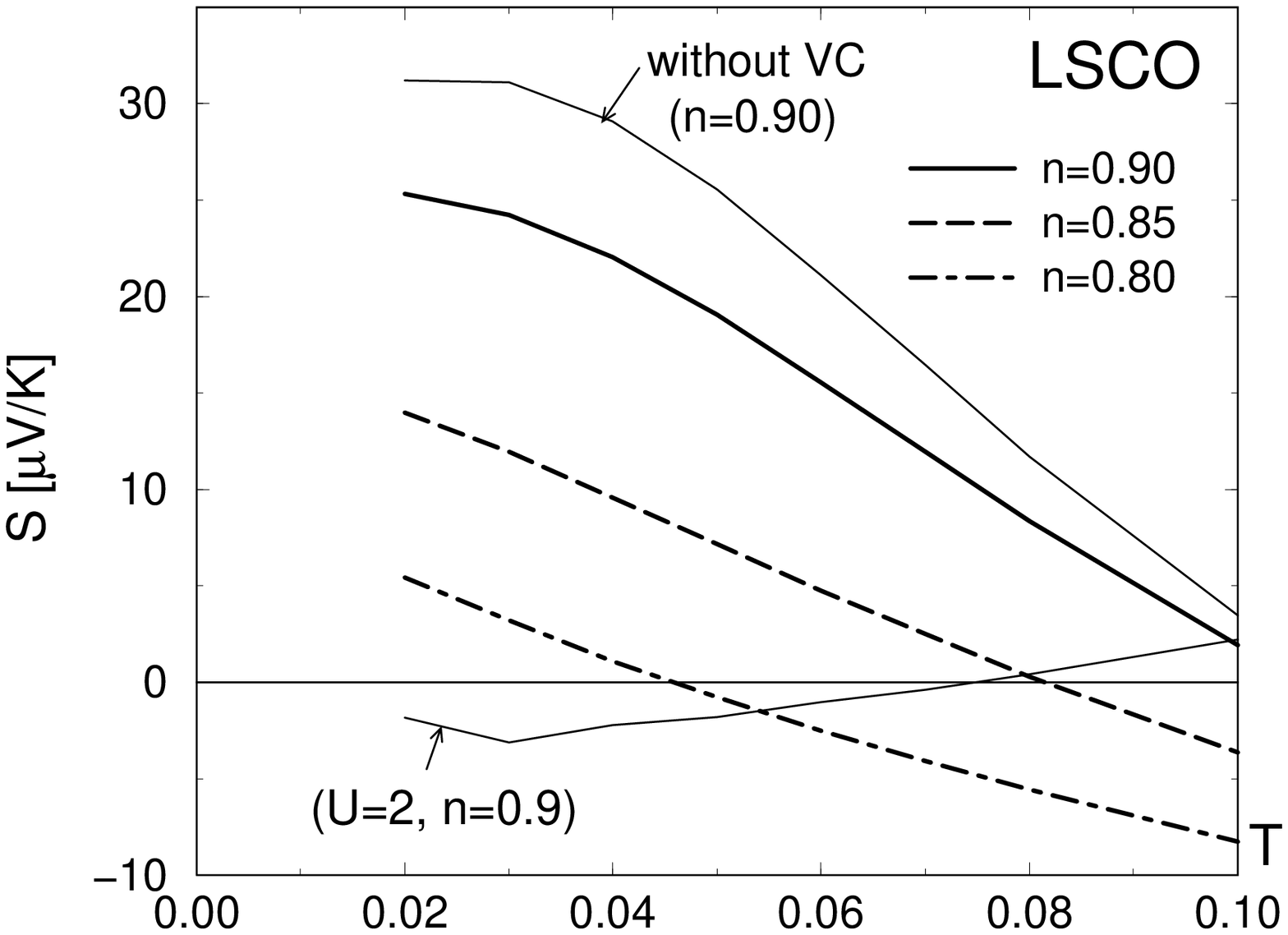,width=6.5cm}
\end{center}
\caption{
Temperature dependences
of the TEP calculated with full VC's for ${\vec J}_\k$. 
Here, $T=0.1$ corresponds to $\sim 500$~K.
We put $U=8$ for YBCO, $U=6$ for NCCO and $U=5.5$ for LSCO.
For NCCO, the VC's play a qualitatively
essential role, and 
$|S(n=1.10)|>|S(n=1.15)|$ is realized only when the VC's are 
taken into account.
In contrast,
the VC's are less important for YBCO and LSCO.
}
 \label{fig:S-T}
\end{figure}
In Fig.~\ref{fig:S-T},
we show numerical results of the TEP derived from eq.~(\ref{eqn:Lxx})
according to the conserving approximation.
The TEP without the VC for $L_{xx}$, which is given by replacing
$J_{\k x}(\e)$ with $v_{\k x}(\e)$ in eq.~(\ref{eqn:Lxx}),
is also plotted in Fig.~\ref{fig:S-T}.
Here, $T=0.02$ corresponds to $\sim100$~K.
In YBCO ($U=8$), $dS/dT<0$ for $n=0.80\sim0.92$,
nd $S$ becomes positive at room temperatures
for optimally- and under-doped case ($n\ge0.85$).
This result is consistent with that of experiments
 \cite{Cooper}.
We stress that $S \approx a\cdot T$ and $a<0$ 
for a smaller interaction ($U=3.5$) 
as shown in Fig.~\ref{fig:S-T},
which indicats the importance of the correlation effect on $S$.

On the other hand, $S$ is always negative 
and $|S|$ increases as $n$ approaches 1 in NCCO, and 
$dS/dT>0$ in under-doped systems for higher temperatures.
These results are consistent with those of experiments
 \cite{Sato,Fournier}.
We stress that the VC for ${\vec J}_\k$ 
enhances the anomalous temperature dependence of $S$ 
strongly in the case of NCCO,
as shown in Fig.~\ref{fig:S-T}.
Finally, $dS/dT<0$ is also reproduced in the case of LSCO,
and $|S|$ for LSCO is larger than $|S|$ for YBCO or NCCO,
if we compare the same filling cases.
However, we see that the qualitative behavior of $S$
is the same if the Fermi surface is hole-like.
We note that $S$ given by eq.~(\ref{eqn:Lxx}) becomes zero at 
$T=0$ if the ground state is metallic.
This finding indicates that $dS/dT$ should change to positive below 
$T=0.02$ for both YBCO and LSCO.

\begin{figure}
\begin{center}
\begin{minipage}{0.45\linewidth}
\epsfig{file=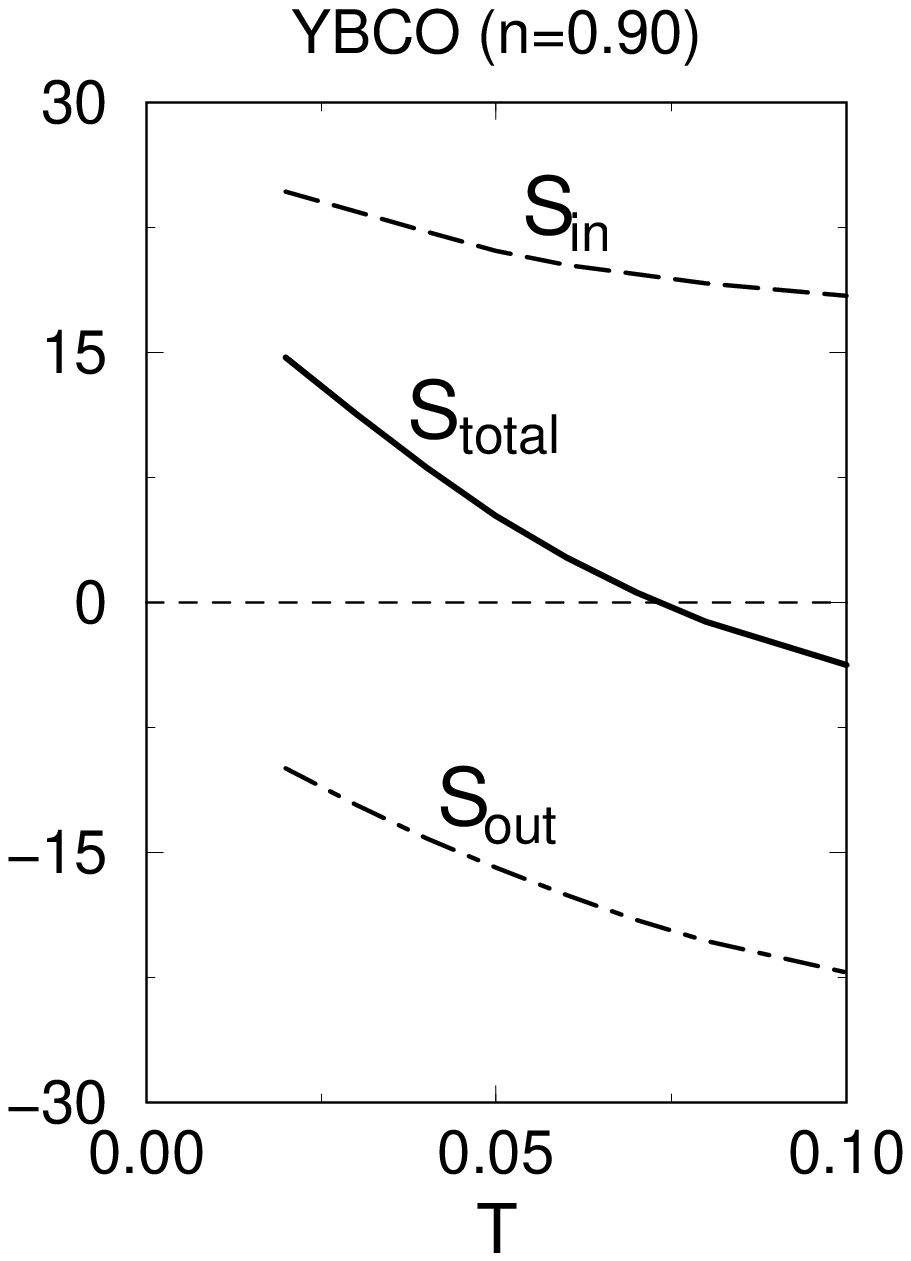,width=3.5cm}
\end{minipage}
\begin{minipage}{0.45\linewidth}
\epsfig{file=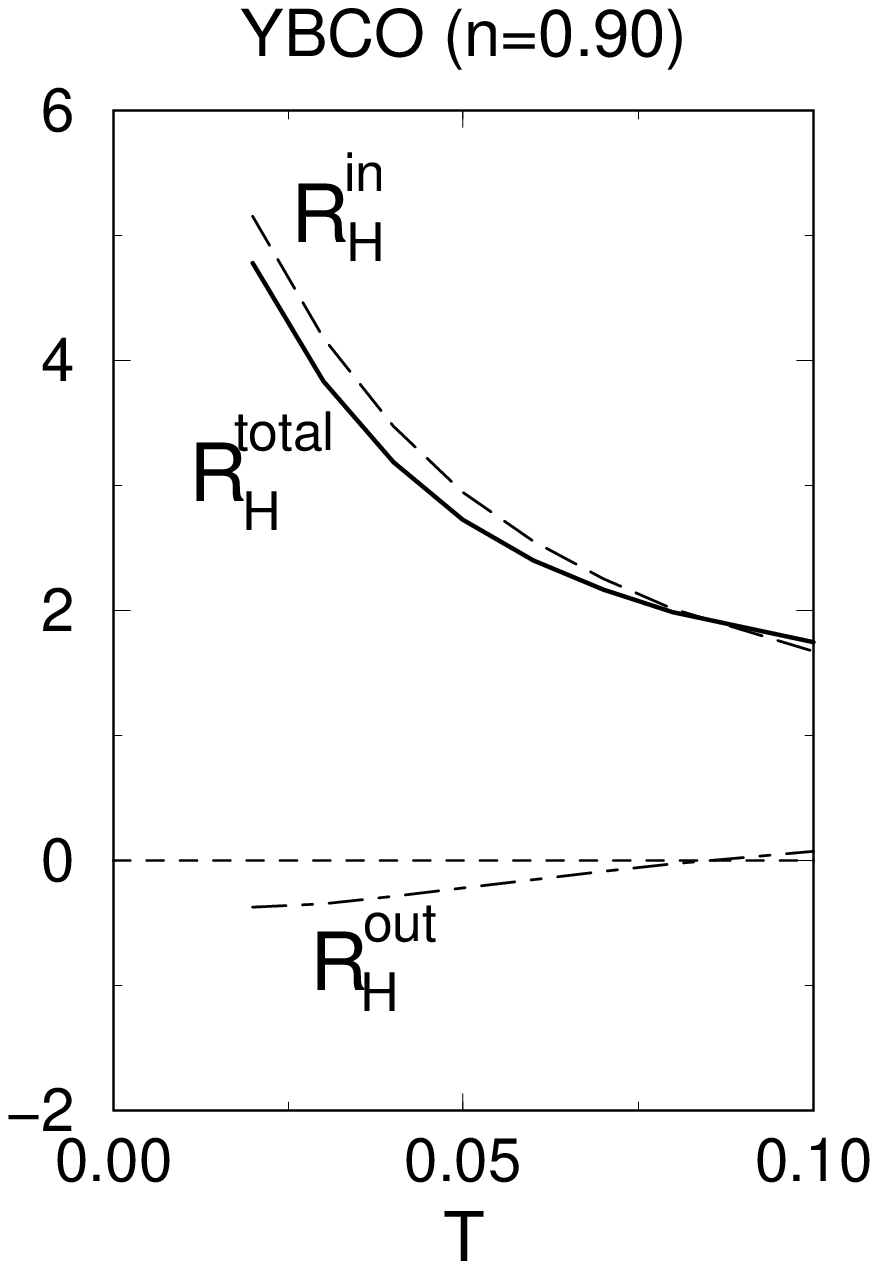,width=3.3cm}
\end{minipage}
\end{center}
\caption{
Temperature dependences of
$S_{\rm in}$ and $S_{\rm out}$, and
$R_{\rm H}^{\rm in}$ and $R_{\rm H}^{\rm out}$ 
for YBCO ($n=0.90$).
Here $S_{\rm total} = S$ and 
$R_{\rm H}^{\rm total} = R_{\rm H}$, respectively.
}
 \label{fig:S-RH-inout}
\end{figure}
Finally, we discuss the following quantity
to understand qualitatively the temperature and doping dependence of
$S$ for YBCO:
\begin{eqnarray}
 L_{xx}^{\rm in} &=& \sum_{\k}^{|k_x|+|k_y|<\pi} 
 \int \frac{d\e}{\pi} \left(-\frac{\d f}{\d\e} \right) 
 \e v_{\k x}(\e) \Bigl[ \ \cdots \ \Bigr],
\end{eqnarray}
where $[ \ \cdots \ ]$ is the same as that of eq.~(\ref{eqn:Lxx}).
In Fig.~\ref{fig:S-RH-inout},
we show $S_{\rm in}\equiv -eL_{xx}^{\rm in}/T\s_{xx}$ together with 
$S_{\rm out}\equiv S-S_{\rm in}$.
Apparently, 
$S_{\rm in[out]}$ represents the contribution from
the inside [outside] region of the AFBZ.
As we expected analytically,
$S_{\rm in}>0$ and $S_{\rm out}<0$ is actually observed, 
and $|S_{\rm in}|>|S_{\rm out}|$ for YBCO at lower temperatures
because the cold spots exist inside the AFBZ.
In Fig.~\ref{fig:S-RH-inout},
we see a conventional behavior
$S_{\rm out}\propto T$ approximately.
On the other hand, 
$S_{\rm in}\sim$const. is realized  because
$1/\gamma_\k(\e_\k^\ast)$ around the cold spot
becomes much asymmetric 
as the temperature decreases,
as shown in Fig.~\ref{fig:del-Ek}.
According to eq.~(\ref{eqn:gamma}),
we see that 
$\d\gamma_\k/\d k_\perp \propto \xi^2 \Delta k \cdot \gamma_\k$ 
for $\xi\Delta k \simle 1$, and
$\d\gamma_\k/\d k_\perp \propto (\Delta k)^{-1} \cdot \gamma_\k$ 
for $\xi\Delta k \simge 1$.
Thus, the $k$-dependence of 
$1/\gamma_\k(\e_\k^\ast)$ near the cold spot
is prominent in the case of 
$\xi\sim(\Delta k_{\rm cold})^{-1}$,
which is realized in YBCO for $n=0.9$ at low temperatures
according to our previous study
 \cite{Kontani}.
Moreover, $S$ is enhanced by $z^{-1}$, which increases slowly
as the temperature decreases. 
As a result, the approximate behavior 
$S (=S_{\rm in}+S_{\rm out}) \propto T+a$ $(a>0)$ as well as the
change of the sign of ${S}$ in YBCO is realized.

This situation is very contrastive to that for 
the Hall coefficient.
We consider $R_{\rm H}^{\rm in}$ which comes from
the inside region of the AFBZ, and 
$R_{\rm H}^{\rm out} \equiv R_{\rm H} - R_{\rm H}^{\rm in}$.
As shown in Fig.~\ref{fig:S-RH-inout},
$|R_{\rm H}^{\rm in}| \gg |R_{\rm H}^{\rm out}|$
for $0.1\ge T \ge 0.02$, which means that
$R_{\rm H}$ is almost determined by the electronic 
property at the cold spot.
As a result, the simple scaling relations 
$R_{\rm H}\sim\xi^2$ and $\Delta\rho\cdot\rho\sim\xi^4$
are realized in high-$T_{\rm c}$ cuprates 
 \cite{Kontani,MR-HTSC}.

In summary,
we analyzed the anomalous behavior of the TEP in 
high-$T_{\rm c}$ cuprates in a conserving manner.
In our numerical calculation based on the spin-fluctuation theory,
the main features of the TEP are reproduced successfully, 
at least for $|1-n|\simge0.1$ above the spin pseudo-gap temperature.
The main origin is that
the quasiparticle damping rate, $\gamma_\k(\e_\k^\ast)$,
becomes more anisotropic near the Fermi surface
as the temperature is decreased, 
reflecting the growth of AF fluctuations.
In conclusion, the {\it seemingly} non-Fermi liquid behavior 
of the TEP is well understood in terms of the  
nearly AF Fermi liquid picture.
In particular, the difference in the sign of the TEP
in electron- and hole-doped systems is naturally explained 
by the "cold spot alternation" mechanism.
Conversely, the success of the present study means that
the AF fluctuation theory provides a reliable description 
for the excited states of quasiparticles,
which determine the thermotransport phenomena.

Before concluding the study,
we comment that compounds 
with large figure of merit, $Z=\s_{xx}S^2/\kappa$
($\kappa$ being the thermal conductivity),
attract great attention nowadays because of their applicability
in electricity generators or refrigerators
 \cite{Mahan}.
Thus, theoretical study of the TEP phenomena 
in strongly correlated systems will become much more important
in the near future.

The author is grateful to T. Saso
for stimulating discussions and comments.
He also thanks K. Yamada for useful comments.


\end{multicols}


\begin{thebibliography}{99}
  
\bibitem{Cooper}
 S.~D. Obertelli, J.R. Cooper, and J.L. Tallon:
 Phys. Rev. B {\bf 46} (1992) 14928.

\bibitem{Sato}
 J. Takeda, T. Nishikawa, and M. Sato: Physica C {\bf 231} (1994) 293.

\bibitem{Fournier}
 P. Fournier, X. Jiang, W. Jiang, S.N. Mao, T. Venkatesan, C.J. Lobb,
 and R.L. Greene: Phys. Rev. B. {\bf 56} (1997) 14149.

\bibitem{ARPES-NCCO}
N.~P. Armitage, D.~H. Lu, D.~L. Feng, C. Kim, A. Damascelli, 
 K.~M. Shen, F. Ronning, and Z.-X. Shen: 
 Phys. Rev. Lett. {\bf 86} (2001) 1126.
\bibitem{ARPES-NCCO2}
 N.P. Arimitage {\it et. al.}: preprint (cond-mat/0107244).

\bibitem{d-inf1} 
 H. Schweitzer and G. Czycholl: Phys. Rev. Lett. {\bf 67} (1991) 3724.
\bibitem{d-inf2} 
 G. P{\' a}lsson and G. Kotliar: Phys. Rev. Lett. {\bf 80} (1998) 4775.

\bibitem{Baym}
 G. Baym and L.P. Kadanoff: Phys. Rev. {\bf 124} (1961) 287.

\bibitem{Kontani} 
 H. Kontani, K. Kanki and K. Ueda:
 Phys. Rev. B {\bf 59} (1999) 14723.
 
\bibitem{Kanki} 
K. Kanki and H. Kontani:
 J. Phys. Soc. Jpn. {\bf 68} (1999) 1614.

\bibitem{Bickers}  
N.E. Bickers and S.R. White: Phys. Rev. B {\bf 43} (1991) 8044.

\bibitem{Dahm} 
 T. Dahm and L. Tewordt: Phys. Rev. B {\bf  52} (1995) 1297.

\bibitem{MR-HTSC}
 H. Kontani: to appear in J. Phys. Soc. Jpn. (2001), No.7
 (cond-mat/0011327).

\bibitem{Luttinger} 
 J.~M. Luttinger: Phys. Rev. {\bf  135} (1964) A1505.

\bibitem{Mahan}
 M. Jonson and G.~D. Mahan: Phys. Rev. B {\bf 42} (1990) 9350.
 
 \bibitem{Future}
 H. Kontani : in preparation.

\bibitem{Eliashberg}
 G.~M. Eliashberg :
 Sov. Phys. JETP {\bf 14} (1962) 886.

\bibitem{Yamada}
  K. Yamada and K. Yosida: Prog. Theor. Phys. {\bf 76} (1986) 621.

\bibitem{MR-formula}
 H. Kontani: Phys. Rev. B {\bf 64} (2001) 054413.

\bibitem{BEDT-Hall}
 H. Kontani and H. Kino: Phys. Rev. B {\bf 63} (2001) 134524.

\bibitem{SCR}
 T. Moriya, Y. Takahashi and K. Ueda:
 J. Phys. Soc. Jpn {\bf 59} (1990) 2905.

\bibitem{Pines}
 B.~P. Stojkovi{\'c} and D. Pines: Phys. Rev. B {\bf 55} (1996) 857.

\end{thebibliography}
\end{document}